\documentclass[final]{svjour2}
\usepackage{graphicx}
\usepackage{rotating}
\usepackage{amssymb}
\usepackage{mathptmx}
\usepackage[numbers]{natbib}
\makeatletter
\journalname{Journal of Low Temperature Physics}

\bibpunct{}{}{,}{s}{}{,}

\begin{document}

\title{Zeeman splitting and nonlinear field-dependence in superfluid $^3$He}
\author{C.A. Collett \and J. Pollanen \and J.I.A. Li \and W.J. Gannon \and W.P. Halperin}
\institute{Department of Physics and Astronomy, Northwestern University, Evanston, IL 60208, USA.\\
\email{ccollett@northwestern.edu}}
\date{\today}

\maketitle

\begin{abstract}
We have studied the acoustic Faraday effect in superfluid $^3$He up to significantly larger magnetic fields than in previous experiments achieving rotations of the polarization of transverse sound as large as 1710$^\circ$. We report nonlinear field effects, and use the linear results to determine the Zeeman splitting of the imaginary squashing mode (ISQ) frequency in $^3$He-$B$.
\end{abstract}

\section{Introduction}
Superfluid $^3$He is a $p$-wave, spin-triplet superfluid of great interest due to its unconventional pairing symmetry. While numerous experimental probes exist, transverse zero sound has recently become an excellent tool to explore the order parameter structure, after its prediction by Landau \cite{lan.57b} and Moores and Sauls \cite{moo.93}, and discovery by Lee \textit{et al.}\cite{lee.99} The coupling between transverse sound and the Imaginary Squashing Mode (ISQ), an order parameter collective mode, allows for spectroscopy of the mode and its behavior in a magnetic field. The ISQ has total angular momentum $J=2$ with five Zeeman sub-states. Transverse sound couples to the $m_J=\pm1$ states, causing a rotation of the sound polarization in a magnetic field, called the Acoustic Faraday Effect (AFE). It is this effect that both proves the existence of transverse zero sound in the superfluid\cite{lee.99} and provides a sensitive probe into the magnetic field dependence of the ISQ.\cite{dav.08a}
\begin{figure}[h!]
\begin{center}
\includegraphics[%
  width=0.75\linewidth,
  keepaspectratio]{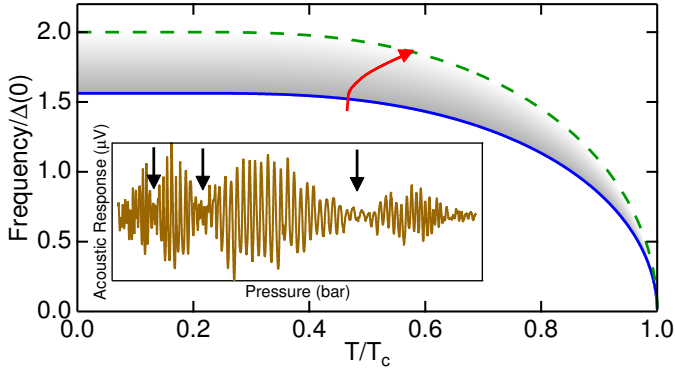}
\end{center}
\caption{\label{fig:EnDiag}(Color online) ISQ (blue line) and pair-breaking (green dashed line) frequencies plotted against reduced temperature. A typical pressure sweep alters the sound frequency normalized to the zero temperature gap to follow the red line, and transverse sound only propagates in the grey shaded area. Inset: An acoustic trace versus pressure at $H=0.04$ T, illustrating the cavity oscillations from changes in the sound velocity and the envelope from the AFE. The black arrows indicate AFE rotations of $90^\circ$, $270^\circ$, and $450^\circ$, from right to left.}
\end{figure}

Transverse sound interacts with the ISQ for acoustic frequencies near the mode according to the dispersion relation,\cite{moo.93}
\begin{equation}
\frac{\omega^2}{q^2v_F^2}=\Lambda_0+\Lambda_{2^-}\frac{\omega^2}{\omega^2-\Omega^2(T,P,H)-\frac{2}{5}q^2v_F^2}, \label{fulldisp}
\end{equation}
where $\omega$ is the sound frequency, $q$ the wavevector, $v_F$ the Fermi velocity, and $\Omega$ the ISQ frequency, with a zero-field value\cite{dav.06,dav.08a} of $\sim\sqrt{12/5}$ times the weak-coupling-plus gap.\cite{rai.76} $\Lambda_0=\frac{F_1^S}{15}(1-\lambda)(1+\frac{F_2^s}{5})/(1+\lambda\frac{F_2^s}{5})$ is the quasiparticle restoring force, and $\Lambda_{2^-}=\frac{2F_1^s}{75}\lambda(1+\frac{F_2^s}{5})^2/(1+\lambda\frac{F_2^s}{5})$ is the superfluid coupling strength, where the $F_n^s$ are Landau parameters, and $\lambda$ is the Tsuneto function.\cite{sau.00a} We model the field dependence of the dispersion relation over a wide range of frequency by simply replacing the denominator on the right-hand side of Eq. \ref{fulldisp} with
\begin{equation}
D^2=\omega^2-\Omega_0^2-\frac{2}{5}q^2v_F^2-m_JA\gamma_{\mbox{eff}}H-B\gamma_{\mbox{eff}}^2H^2-m_JC\gamma_{\mbox{eff}}^3H^3. \label{ABC}
\end{equation}
Here $\Omega_0$ is the zero-field ISQ frequency, $\gamma_{\mbox{eff}}$ is the effective gyromagnetic ratio of $^3$He, $H$ is the external magnetic field, and the terms containing $A$, $B$, and $C$ describe the linear, quadratic, and cubic magnetic field dependence of the dispersion, respectively. The $m_J$ dependence of each term is simplified to reflect the fact that $m_J=\pm1$, making the linear and cubic terms either positive or negative but having no effect on the quadratic term.

\section{Experiments}
In these experiments we use a dilution refrigerator and demagnetization stage setup described previously\cite{dav.08a} to reach $\sim600$ $\mu$K with a $^3$He cavity formed by a transducer and a reflecting plate. To probe the Faraday rotation, we lower the pressure of the helium from $\sim6$ to 3 bar, changing the frequencies of the ISQ and pair-breaking, which transform relative to the acoustic frequency of 88 MHz as shown by the trajectory of the red arrow in Fig. \ref{fig:EnDiag}, taking into account heating inherent in the experiment which raises the temperature from $\sim0.5\to0.6T_c$. This causes changes in $D^2$ and thus the sound velocity, $c_t$, leading to oscillations in the transducer response, or cavity oscillations, such as those seen in the inset to Fig. \ref{fig:EnDiag}. With a magnetic field present, the polarization rotation is seen as the overall envelope in the inset with minima indicated by the black arrows. These effects are governed by
\begin{equation}
V_Z\propto \cos\theta\sin\Big(\frac{2d\omega}{c_t}\Big), \label{transig}
\end{equation}
where $V_Z$ is the transducer voltage, $\theta$ is the AFE rotation angle, and $d=31.6\pm0.1 \mu$m is the cavity spacing.\cite{dav.08a}

Using Eq. \ref{transig}, we can get both $\theta$ and $c_t$ from our acoustic response data. By measuring the envelope we calculate $\theta$ based on its amplitude, using the nodes as fixed angles $\theta=n\times90^\circ;\:n=1,3,5\dots$. We also convert the period of the cavity oscillations to changes in the sound velocity using
\begin{equation}
1 \mbox{ Period}=2d\frac{\omega}{2\pi}\Big|\frac{1}{c_{tf}}-\frac{1}{c_{ti}}\Big|. \label{dcper}
\end{equation}
After calculating an initial value of $c_t$ near the ISQ, we use the sound velocity changes calculated from Eq. \ref{dcper} to extend that initial sound velocity over the entire data range.

\begin{figure*}[t!]
\begin{center}
\includegraphics[%
  width=0.75\linewidth,
  keepaspectratio]{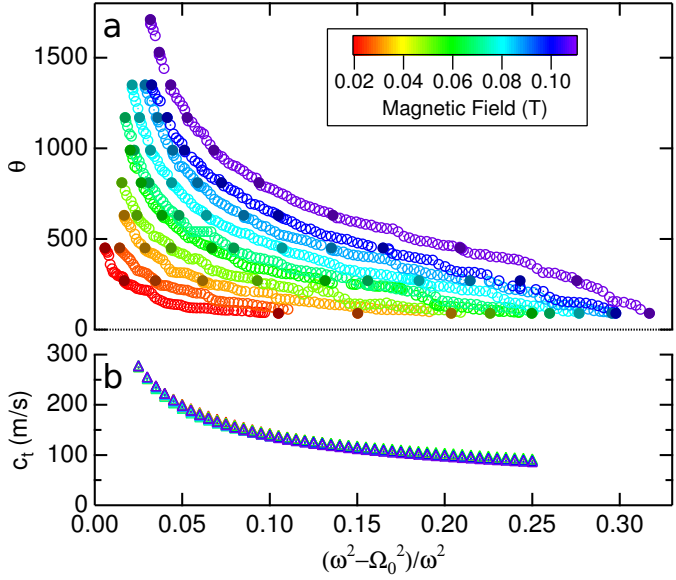}
\end{center}
\caption{\label{fig:data}(Color online) Data taken from acoustic traces at all experimental fields, plotted against the shift: (a) Sound polarization rotation angle data, $\theta$. Open circles are taken from the envelope amplitude, while solid circles correspond to minima in the acoustic traces. (b) Sound velocity data, $c_t$. The data points at each shift mostly overlap.}
\end{figure*}

\section{Results and Discussion}
We show our data in Fig. \ref{fig:data}, plotted against the normalized difference in the square of the frequencies, $(\omega^2-\Omega_0^2)/\omega^2$. This scaling, which we refer to as the frequency shift, or just the shift, better reflects the changes in the dispersion caused by the changes in temperature and pressure during an experimental run. $\theta$ shows strong field dependence, but $c_t$ appears to be nearly field independent at these fields. We relate this data to the dispersion by taking Eq. \ref{fulldisp}, with the right-hand side denominator given by Eq. \ref{ABC}, and using the following definitions:\cite{sau.00a,sau.00b}
\begin{equation}
\theta=2d\,\delta q, \label{theta}
\end{equation}
\begin{equation}
c_t=2\omega/(q_++q_-), \label{ct}
\end{equation}
where $\delta q=|q_+-q_-|/2$, and $q_\pm$ is obtained by solving Eq. \ref{fulldisp} for $q$, setting $m_J=\pm1$. Using Eqs. \ref{theta} and \ref{ct} we find that $c_t$ depends most strongly on the quadratic field term, while $\theta$ depends on the linear and cubic terms, due to the cancelling effects of $m_J=\pm1$.
\begin{figure}[t!]
\begin{center}
\includegraphics[%
  width=0.55\linewidth,
  keepaspectratio]{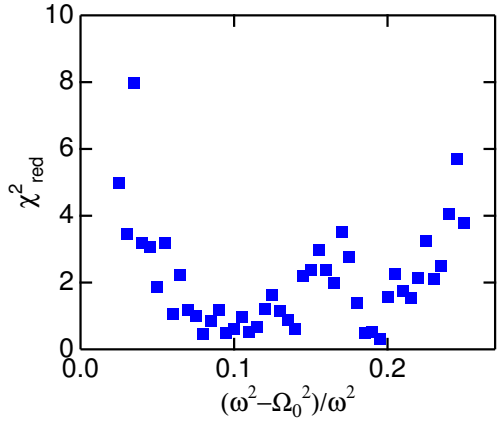}
\end{center}
\caption{\label{fig:chired}(Color online) Values of $\chi_{\mbox{red}}^2$ calculated using Eq. \ref{chi2red} for fitting $\theta$ to the dispersion using $A$ and $C$ as free parameters, with $B$ as the upper bound from fitting $c_t$.}
\end{figure}

As mentioned above, $c_t$ shows very little field dependence. In order to quantify this we group the data in frequency shift bins spaced by $(\omega^2-\Omega_0^2)/\omega^2=0.005$, and fit each bin to Eq. \ref{fulldisp} as a function of field, using Eq. \ref{ABC} with $B$ as the only free parameter and setting $A=C=0$. All fits are done using the Levenberg-Marquardt algorithm\cite{lev.44,mar.63} for nonlinear least-squares minimization. We evaluate these fits quantitatively by calculating $\chi_{\mbox{red}}^2$, a measure of how well a model fits data with a normal sample distribution, given by
\begin{equation}
\chi_{\mbox{red}}^2=\frac{1}{\nu}\sum_i^N{\frac{(y_i-f_i)^2}{\sigma_i^2}}, \label{chi2red}
\end{equation}
where $\nu=N-M$ is the number of degrees of freedom, $N$ is the number of data points, $M$ is the number of fit parameters, $y_i$ is the measured value, $f_i$ is the calculated value, and $\sigma_i$ is the standard deviation of the measurement. For $c_t$ we calculate $\sigma_i$ directly from the data, assuming zero field dependence, resulting in values of $\sigma\sim2$ m/s, while for $\theta$ we estimate $\sigma=10^\circ$ from our measurements. For a fit either $\sigma$ or $2\sigma$ from our data, $\chi_{\mbox{red}}^2\sim1.25$ or 5.5, respectively.

\begin{figure}[t!]
\begin{center}
\includegraphics[%
  width=0.75\linewidth,
  keepaspectratio]{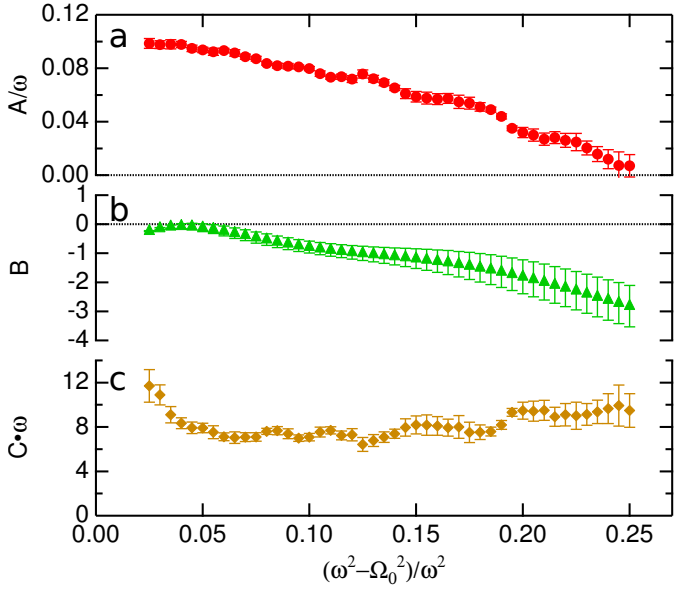}
\end{center}
\caption{\label{fig:ABC}(Color online) $A$, $B$, and $C$ parameters for Eq. \ref{ABC}, made dimensionless by normalizing to appropriate orders of $\omega$. (a) The linear field dependence $A/\omega$, red circles. (b) An upper bound on the quadratic field dependence $B$, green triangles. (c) The cubic field dependence, $C\cdot\omega$, solid orange diamonds. The upturn near $\omega=\Omega_0$ is in a region of greater uncertainty due to the lack of the highest field data, as described in the text.}
\end{figure}

The results of our $c_t$ fits only give us an upper bound on its quadratic field dependence, which constrains the $B$ term such that it has little to no effect on $\theta$. Consequently, to account for the non-linear dependence of $\theta$ when we fit $\theta$ to Eq. \ref{fulldisp} as above, we must include the cubic $C$ term in Eq. \ref{ABC} as well as the $A$ term as free parameters, using the $B$ from $c_t$ as an input, giving an accurate fit to the data resulting in the $\chi_{\mbox{red}}^2$ values shown in Fig. \ref{fig:chired}. At the lowest and highest shifts $\chi_{\mbox{red}}^2$ increases, which is reflected in larger error bars for $A$ and $C$ at these shifts. We attribute these larger error bars to a larger uncertainty in $\theta$ due to a greater difficulty in precisely identifying the rotation angle in these regions. Our $A$, $B$, and $C$ results are shown in Fig. \ref{fig:ABC}, given in a dimensionless form by normalizing to appropriate orders of $\omega$.

The appearance of an upturn at low shifts in $C$ can be attributed to the lack of high field $\theta$ data at those shifts, due to the rotation angle changing too fast to be determined. At higher shifts, $C$ is a fairly constant value. Our data thus shows a decreasing linear term, an increasing quadratic bound, and a constant cubic term with increasing shift.

Although there are predictions \cite{sch.83} for the quadratic field dependence of the ISQ, our $B$ values cannot be directly compared with them due to the insensitivity of $c_t$ to the magnetic field. As our data provides only an upper bound on the magnitude of the quadratic term in the dispersion we can only say that previous measurements\cite{mov.88} seem to fall within that bound at the mode. There are currently no theoretical predictions for the cubic field dependence of the ISQ.

\begin{figure}[t!]
\begin{center}
\includegraphics[%
  width=0.75\linewidth,
  keepaspectratio]{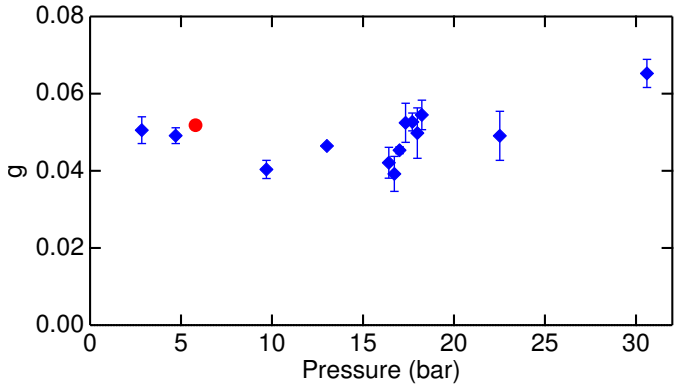}
\end{center}
\caption{\label{fig:gs}(Color online) Land\'e $g$-factor values as a function of pressure, for our data (red circle, error bars inside data point) and reanalyzed data from Davis \textit{et al.} (blue diamonds).\cite{dav.08a,col.12}}
\end{figure}

By extrapolating our $A$ data to $\omega=\Omega_0$, we can apply the theory of Sauls and Serene\cite{sau.82} to extract the Land\'e $g$-factor of the ISQ. The $g$-factor is a measure of the magnitude of the linear Zeeman splitting of the ISQ, and is predicted \cite{sau.82} to depend on the magnitude of $f$-wave pairing interactions in $^3$He. As the theory applies only in the region around $\omega\sim\Omega$, we cannot extract more than a single data point, which, using the relationship $A_0=2g\omega$, where $A_0$ is the extrapolated value of $A$ to $\omega=\Omega_0$, is shown as a red circle in Fig. \ref{fig:gs}. We also present reanalyzed data from Davis \textit{et al.} as blue diamonds.\cite{dav.08a} The reanalysis, which focuses on their extrapolation of their data to $T=0$ rather than to $\omega=\Omega_0$, as well as a discussion of the applicability of the theory, will be discussed in a forthcoming publication.\cite{col.12}

\section{Conclusion}
We have measured the splitting of the ISQ using transverse sound, finding that nonlinear field effects play a significant role at fields up to $H=0.11$ T. Using a simple model for the effect of splitting on the dispersion we have quantified the field dependence up to cubic order. We have set an upper bound on the quadratic splitting, which is small at these fields, and have determined the linear and cubic splitting parameters, as well as the Land\'e $g$-factor of the ISQ.

\section{Acknowledgments}
We would like to thank J.A. Sauls for his help with this work, and acknowledge the support of the National Science Foundation, DMR-1103625.

\bibliographystyle{apsrev4-1}
\bibliography{refs}{}
\end{document}